\documentclass[12pt]{iopart}

\usepackage{graphicx}
\usepackage{dcolumn}
\usepackage{bm}
\usepackage{color}
\usepackage{ulem}
\usepackage[utf8]{inputenc}
\usepackage{siunitx}
\usepackage{hyperref}

\newcommand{\Qi}{Q_{\mathrm{i}}}
\newcommand{\Qe}{Q_{\mathrm{e}}}

\begin{document}

\title[Thin-film surface acoustic wave resonators]{Surface acoustic wave resonators on thin film piezoelectric substrates in the quantum regime}

\author{Thomas Luschmann$^{1,2,3}$, Alexander Jung$^{1,2}$, Stephan Geprägs$^{1}$, Franz X. Haslbeck$^{1,2,3}$, Achim Marx$^{1}$, Stefan Filipp$^{1,2,3}$, Simon Gröblacher$^{4}$, Rudolf Gross$^{1,2,3}$ and Hans Huebl$^{1,2,3}$}
\address{$^1$ Walther‑Meißner‑Institut, Bayerische Akademie der Wissenschaften, Walther‑Meißner‑Str.8, 85748 Garching,
Germany}
\address{$^2$ Technische Universität München, TUM School of Natural Sciences, Physics Department, James‑Franck‑Str.1, 85748 Garching, Germany}
\address{$^3$ Munich Center for Quantum Science and Technology (MCQST), Schellingstr.4, 80799 Munich,
Germany}
\address{$^4$ Kavli Institute of Nanoscience, Department of Quantum Nanoscience, Delft University of Technology, 2628CJ Delft, The Netherlands}
\eads{\mailto{thomas.luschmann@wmi.badw.de}, \mailto{hans.huebl@wmi.badw.de}}
\vspace{10pt}
\begin{indented}
\item[]January 2023
\end{indented}

\begin{abstract}
Lithium Niobate (LNO) is a well established material for Surface Acoustic Wave (SAW) devices including resonators, delay lines and filters. Recently, multi-layer substrates based on LNO thin films have become commercially available. Here, we present a systematic low-temperature study of the performance of SAW devices fabricated on LNO-on-Insulator (LNOI) and LNO-on-Silicon substrates and compare them to bulk LNO devices. Our study aims at assessing the performance of these substrates for quantum acoustics, i.e.\ the integration with superconducting circuits operating in the quantum regime. To this end, we design surface acoustic wave resonators with a target frequency of $\SI{5}{GHz}$ and perform experiments at millikelvin temperatures and microwave power levels corresponding to single photons or phonons. The devices are investigated regarding their internal quality factors as a function of the excitation power and temperature, which allows us to characterize and quantify losses and identify the dominating loss mechanism. For the measured devices, fitting the experimental data shows that the quality factors are limited by the coupling of the resonator to a bath of two-level-systems (TLS). Our results suggest that SAW devices on thin film LNO on silicon have comparable performance to devices on bulk LNO and are viable for use in SAW-based quantum acoustic devices.
\end{abstract}

\vspace{2pc}

\maketitle

\section{Introduction}
Surface acoustic waves (SAWs) are highly versatile with applications ranging from sensing (air pressure, magnetic resonance) \cite{lange2008}, signal processing (GHz filters for mobile communications) \cite{ruppel2017} to the sorting of bio-molecules \cite{franke2010}. Recently, an additional research direction has been added to this list: quantum applications. In this field, quantized elastic and acoustic excitations are discussed for information storage due to their expected long lifetimes \cite{oconnell2010,chu2017,chu2018,hann2019,satzinger2018,bienfait2019}. Furthermore, novel interaction regimes are explored in the giant atom limit \cite{friskkockum2014,aref2016a,andersson2019,guo2020}, where the wavelength of the interacting wave becomes comparable to the dimensions of the (artificial) atom. These perspectives inspired the sub-field called circuit quantum acousto-dynamics (cQAD) \cite{manenti2017}. In addition, within cQAD, applications such as quantum signal filters or microwave-to-optics conversion schemes based on the electro-optic properties of piezoelectric materials are considered \cite{tadesse2014,balram2016,shen2020,xu2021}. Other applications are related to superconducting quantum circuits and in particular quantum bits which require the efficient transduction of electromagnetic to elastic excitations and vice versa. However, while materials with large piezo-electric coupling rates are desired as substrates for high performance SAW devices, the same piezo-electric coupling constitutes a critical loss channel limiting the performance of superconducting qubits below state-of-the-art values achieved for devices fabricated e.g.\ on silicon \cite{manenti2017,moores2018}. One approach to overcome these ostensibly conflicting conditions is to physically separate the microwave quantum circuit from the acoustic component of the circuit by putting them on two different substrates and mediate the coupling e.g.\ using a flip-chip assembly in combination with inductive couplers \cite{satzinger2018}. \\
Alternatively, the recent advent of commercially available thin film lithium niobate on insulator (LNOI) wafers enables the exploration of additional routes e.g.\ the planar integration of microwave and mechanical elements on the same chip in a spatially separated fashion. The present LNOI platforms are fabricated in an "ion-slicing"-enhanced wafer-bonding process \cite{poberaj2012} and have been adopted quickly by the integrated photonics community \cite{boes2018}, realizing waveguides \cite{wang2018a}, cavities \cite{cai2016}, as well as electro-optical \cite{guarino2007} and opto-mechanical \cite{jiang2019a} interfaces. Moreover, thin film LNO-on-silicon (without a SiO$_\mathrm{x}$ buffer layer) has recently become a subject of interest with respect to suspended SAW systems \cite{pop2017, sarabalis2020} as well as microwave-to-optical frequency transduction \cite{jiang2020} and investigations of mechanical systems in the quantum regime \cite{arrangoiz-arriola2019}. Previous investigations into the behavior and performance of SAW devices based on thin film piezo-electric platforms proved their general viability, but found considerable differences with respect to their bulk counterparts, which need to be considered in the context of potential applications \cite{kadota2011,hayashi2017,kimura2019}. In particular, a comprehensive study of the performance and loss mechanisms of SAW devices on thin film lithium niobate in the quantum regime, evaluating their viability for applications in quantum acoustics, is still missing.

In this work, we investigate the suitability and performance of LNOI and thin film LNO-on-silicon platforms for SAW-based applications in quantum acoustics. To this end we quantitatively analyze the performance of one of the key building blocks for cQAD, namely GHz-frequency SAW resonators, at conditions typical for the operation in superconducting cQAD architectures. Specifically, this requires studying their properties at millikelvin temperatures and at an average signal power corresponding to the single microwave-frequency photon/phonon level. As a reference point to established SAW technology, we also study devices fabricated on bulk LNO under the same conditions. The fabrication procedure is kept identical between the different materials and we deliberately use a simple and widely established SAW resonator design to make our findings as general and applicable as possible.

\section{Implementation of surface acoustic wave resonators on multi-layer substrates}
\begin{figure}
	 \center{\includegraphics{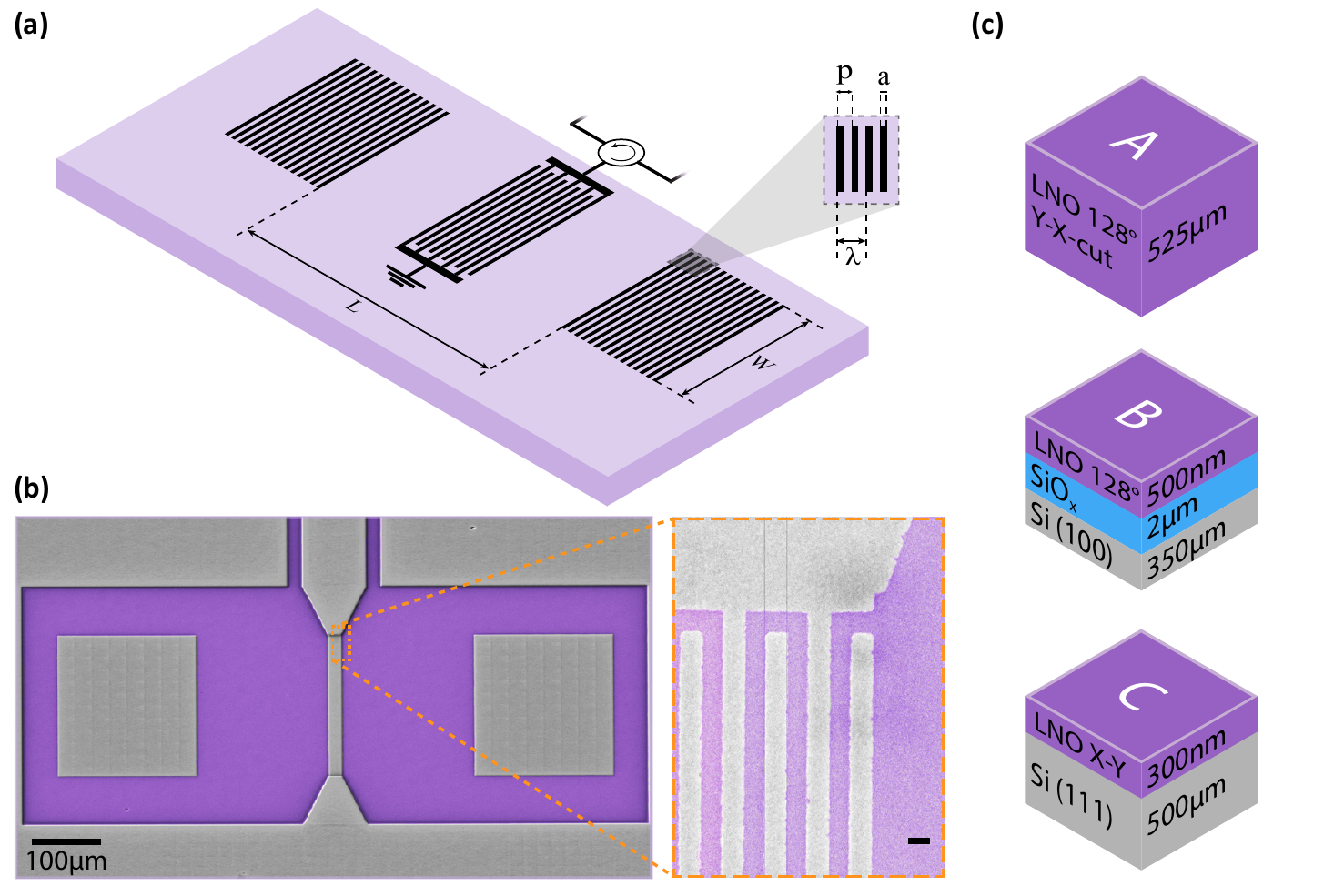}}
	  \caption{\label{fig:intro}(a) Schematic illustration of the surface acoustic wave resonator and the relevant geometric parameters. (b) False color optical micrograph of a SAW resonator structure. Aluminium is shown as gray, while LNO is colored violet. The zoom-in shows a scanning electron micrograph (SEM) highlighting the individual strips of the transducer. The black scalebar corresponds to $\SI{200}{nm}$, the nominal width of the strips. (c) Illustration of the multi-layer structures of the sample types A, B and C studied within the scope of this work.}
	\end{figure}
Surface acoustic wave resonators can be understood as the elastic-wave analogon of an electromagnetic Fabry-Perot cavity, where a standing surface acoustic wave is confined between two highly reflective mirrors [see Fig.\,\ref{fig:intro} (a)]. The mirrors are realized in the form of distributed Bragg reflectors formed by a large number of periodically arranged parallel or curved lines, which modulate the surface mass loading and/or the piezo-electric boundary conditions \cite{morganSAWFilters}. Here, the pitch $p$ and the line width $a$ define the reflective acoustic properties which are characterized as a so-called stop-band of the mirror with center frequency $f_\mathrm{0}$ and bandwidth $\Delta f$ \cite{datta1986}. In our samples, we implement the Bragg mirrors by patterning $N_\mathrm{g}$ straight electrically floating superconducting aluminum metal strips aligned in a parallel fashion on the surface of piezoelectric lithium niobate, as illustrated in Fig.\,\ref{fig:intro} (a). Each strip causes a partial reflection $r$ of the incident surface acoustic wave. At $f_\mathrm{0}$ these partially reflected waves interfere constructively enabling the reflective behavior of the mirror with a total reflection coefficient $\Gamma \approx \tanh(N_g r)$. $\Gamma$ approaches unity for $N_\mathrm{g} r \gg 1$ within the stop-band bandwidth $\Delta f = 2f_\mathrm{0} r/\pi$ \cite{morganSAWFilters}. Two of these mirrors separated by the distance $L$ form a SAW cavity. The resonator supports multiple standing SAW modes within $\Delta f$ separated in frequency by the so-called free spectral range (FSR) $\Delta f_{\textrm{FSR}}=v_{\textrm{p}}/2L_{\textrm{eff}}$, where $v_{\textrm{p}}$ is the phase velocity of the mode and $L_{\textrm{eff}}$ is the effective mirror separation which differs from $L$ due to the distributed reflection inherent to Bragg mirrors \cite{morganSAWFilters, datta1986}. 

To compare the properties of SAW resonators fabricated on standard bulk LNO crystals with devices based on multi-layer substrates, we use three different sample types A, B and C [see Fig.\,\ref{fig:intro} (c)]: (A) A $128^\circ$-rotated Y-X-cut bulk lithium niobate (LNO) crystal - the standard material for SAW resonators and filters \cite{morganSAWFilters,ruppel2017}, (B) a thin film lithium niobate on insulator (LNOI) stack composed of a $\SI{500}{\nano\m}$ thin $128^\circ$-rotated Y-cut LNO thin film on a $\SI{2}{\micro\m}$ thick $\mathrm{SiO_x}$ buffer layer on a $\SI{350}{\micro\m}$ thick Si(100) substrate, and (C) a $\SI{300}{\nano\m}$ thin Y-X cut LNO thin film on a $\SI{500}{\micro\m}$ thick Si(111) substrate.

Superconducting quantum circuits operate naturally at frequencies of several GHz. Thus, their integration with acoustics requires SAW devices operating at these frequencies as well as an efficient transduction of electromagnetic into elastic excitations and vice versa \cite{ekstrom2017}. Our SAW resonators are designed for a center frequency of $f_\mathrm{0}\approx \SI{5}{GHz}$ and we chose standard, single-electrode interdigitated transducers (IDT) for the transduction \cite{morganSAWFilters}. The transducer consists of alternating signal and ground aluminium electrodes [see Fig.\,\ref{fig:intro}\,(a) and (b)] with a pitch of $p=\SI{0.4}{\micro\m}$ and a metallization ratio of $\eta = 0.5$ (i.e. the width of electrodes is equal to the gap in between them). The number of transducer electrodes $N_\mathrm{f}$ is varied between the different devices, but generally kept small in order to avoid significant reflections at the IDT. Therefore the bandwidth of the IDT is much larger than the mirror stop band and the transduction efficiency can be considered constant across the narrow frequency band.

For the mirrors, we use $N_\mathrm{g} = 500$ aluminium strips with a fixed pitch $p=\SI{0.4}{\micro\m}$ for all of our SAW mirrors. The pitch defines the SAW wavelength $\lambda = 2p$, which is reflected most effectively and hereby the operating frequency of the SAW resonator $f_\mathrm{0} = v_{\textrm{p}}/\lambda$. Note that the phase velocity $v_{\textrm{p}}$ can vary between the different layer stacks as the surface wave has an evanescent character with a characteristic penetration depth and thus the elastic properties of the underlying layer can contribute to $v_{\textrm{p}}$ \cite{farnell1972}. For frequencies of about $\SI{5}{GHz}$ the penetration depth of a SAW is approximately $\SI{320}{nm}$ and therefore we expect and observe only slight modifications of $f_\mathrm{0}$. We want to point out that the thickness of the thin films is chosen to allow to maintain the majority of the modes in the lithium niobate and thereby to realize a high effective elastic mode filling factor. 
Finally, it should be noted that different types of surface acoustic waves, including Rayleigh-, shear- and Love-waves, can be excited and confined using this resonator geometry, each with their characteristic phase velocity $v_{\textrm{p}}$ and corresponding resonance frequency \cite{morganSAWFilters}. Moreover, when considering a multi-layer material system, so-called Sezawa modes need to be considered \cite{datta1986}. However, all our devices are designed to excite and confine Rayleigh waves. Therefore, all data shown in the following pertains to Rayleigh-type waves.

All aluminium structures have a thickness of $\SI{20}{nm}$ and are patterned in a single step process using conventional electron beam lithography techniques, electron beam evaporation under UHV conditions, followed by a lift-off process. A false color optical micrograph of a fabricated structure is displayed in Fig.\,\ref{fig:intro} (b).
\begin{figure}
	 \center{\includegraphics{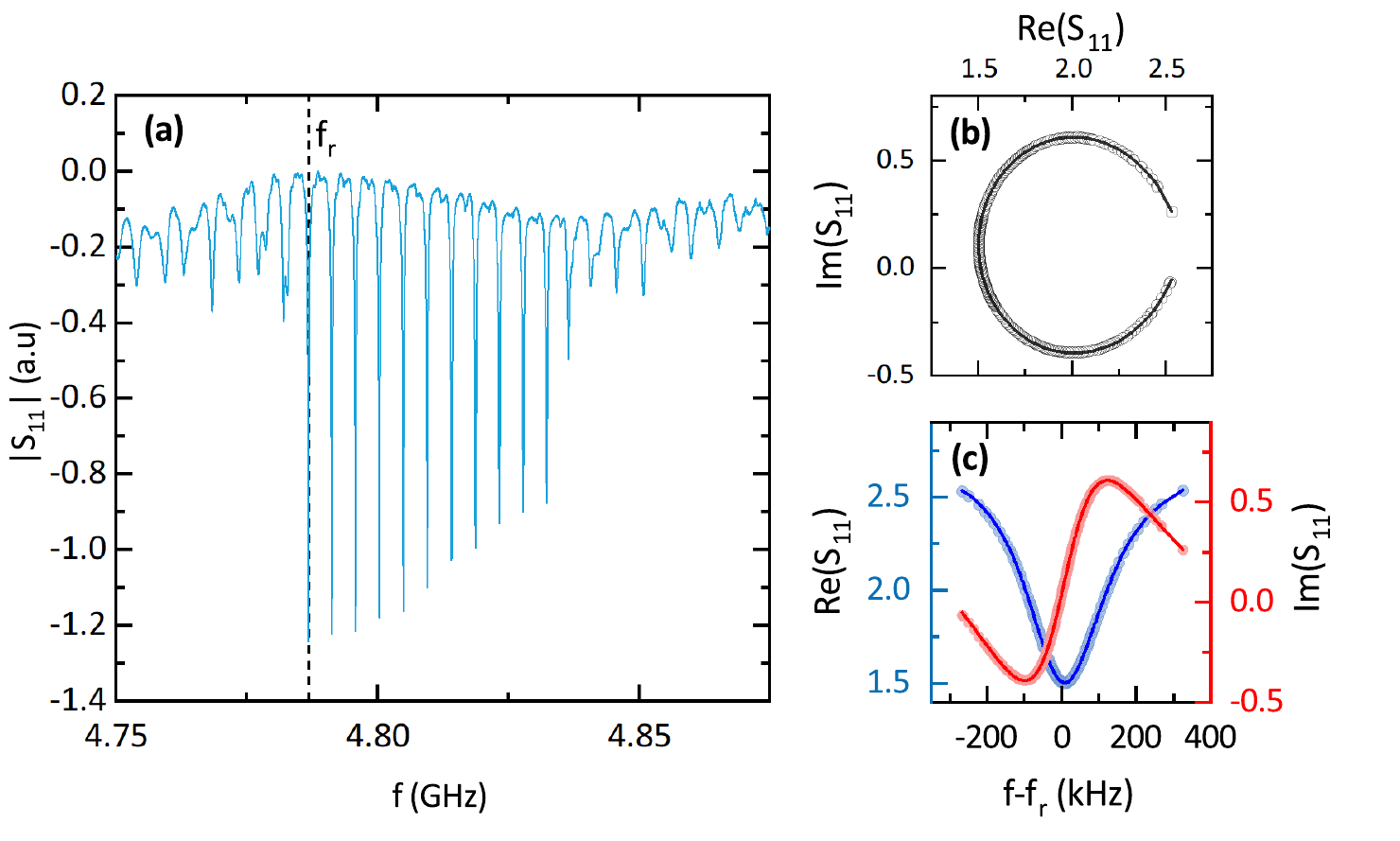}}
	  \caption{\label{fig:circlefit}(a) Normalized amplitude of the reflection spectrum $|S_{11}|(f)$ for a SAW resonator fabricated on the LNO/Si multi-layer stack [sample type C illustrated in Fig.\,\ref{fig:intro} (c)] measured at $T=\SI{20}{mK}$. We identify a single resonance with frequency $f_\mathrm{r}$ marked with a vertical dashed line in the spectrum and fit to Eq.\,(\ref{eq:fit}) in the complex plane (b). (c) Real and imaginary part of $S_{11}$ close to resonance as a function of the relative frequency $f-f_\mathrm{r}$ with their respective fits (solid lines).} 
\end{figure}

\section{Results and Discussion}
\begin{table}
\begin{center}
\begin{tabular}{c c c c c c c c c} \hline \hline
    device & L (\si{mm}) & orientation & r (\%) & W(\si{\mu m}) &  $N_\mathrm{f}$ & $\bar{Q_\mathrm{e}} (10^{3})$ & $Q_\textrm{i} (10^{3}) $ & $Q_\mathrm{i,TLS} (10^{3})$ \\ \hline
    A1 & 0.7 & $90^\circ$ & 1.6 & 391 & 11 &$62.5\pm{12.5}$ & $24.0\pm{0.5}$ & $61.2\pm{1.7}$ \\ 
    A2 & 0.4 & $90^\circ$ & 1.6 & 140 & 31 &$71.5\pm{16.2}$ & $44.4\pm{0.5}$ & $154.7\pm{4.3}$\\
    A3 & 0.3 & $90^\circ$ & 1.9 & 258 & 17 & $39.6\pm{1.8}$ & $49.8\pm{0.5}$ & $234.8 \pm{7.1}$\\
    A4 & 0.2 & $90^\circ$ & 2.0 & 406 & 11 & $35.8\pm{3.3}$ & $64.6\pm{0.6}$ & $270 \pm{6.8}$\\  \hline
    B1 & 0.7 & $0^\circ$ & 0.7 & 285 & 51 & $14.6\pm{2.6}$ & - & - \\
    B3 & 0.3 & $90^\circ$ & 0.6 & 110 & 151 & $18.6\pm{1.4}$ & $2.8\pm{0.1}$ &  $3.3 \pm{0.1}$\\  \hline
    C1 & 0.8 & $90^\circ$ & 0.9 & 502 & 21 & $83.4\pm{9.4}$ & $14.2\pm{6.4}$ & $36.2\pm{0.2}\dagger$ \\
    C2 & 1.2 & $90^\circ$ & 0.7 & 388.5 & 27 & $96.0\pm{1.4}$ & $14.0\pm{12.7}$ & $14.1\pm{7.6}$ \\ 
    C3 & 0.8 & $0^\circ$ & 0.4 & 388.5 & 27 & $35.9\pm{5.1}$ & $9.4\pm{1.7}$ & $35.5\pm{12.3}$ \\
    C4 & 1.2 & $0^\circ$ & 0.3 & 388.5 & 27 & $63.3\pm{0.7}$ & $9.7\pm{1.2}$ & $35.7\pm{7.8}$ \\
\end{tabular}
\caption{Design parameters and experimentally determined properties of the investigated SAW resonator devices. Devices are labeled by sample types A, B and C as illustrated in Fig.\,\ref{fig:intro} (c). All devices operate at $f_\mathrm{0} \approx \SI{5}{GHz}$. The values listed for $Q_\textrm{i}$ correspond to the low temperature ($T_\mathrm{base} \approx \SI{20}{mK}$) and low power (resonator phonon occupation $n_\mathrm{ph} < 1$) limits. The orientation refers to the in-plane propagation direction of SAWs within the resonators w.r.t.~the standard propagation direction of the respective cut (i.e.\ for $128^\circ$-rotated Y-X-cut LNO, $0^\circ$ corresponds to X-propagation). Device B3 could not be measured in the low power level, thus the missing values (-). The given values for $Q_\mathrm{i,TLS}$ are results from fits to Eq.\,(\ref{eq:Qi_TLS_model}), except for the value marked with $\dagger$, which results from Eq.\,(\ref{eq:fitfr}).}
\label{tab:samples}
\end{center}
\end{table}
The SAW resonators are investigated by microwave spectroscopy in a dilution refrigerator with a base temperature of $T_\mathrm{base} \approx \SI{20}{mK}$. Specifically, we measure the complex microwave reflection amplitude $S_{11}(f)$ using a vector network analyzer (VNA) (cf.\ Ref.\,\cite{wang2021}). The microwave input signal is attenuated at the different temperature stages with a total of $\SI{66}{dB}$ attenuation which allows to probe our SAW resonators at the single excitation limit (as discussed in more detail below). The reflected signal is routed through multiple cryogenic circulators, a low-noise cryogenic, and a low noise room-temperature preamplifier to a vector network analyzer.

Figure \ref{fig:circlefit}\,(a), where we plot the normalized magnitude $|S_{11}(f)|$, shows a typical reflection spectrum of a SAW resonator. The multiple narrow and equally spaced absorption features are identified as the standing SAW resonator modes with their characteristic frequency spacing given by the free spectral range $\Delta f_\mathrm{FSR}$. We analyze each SAW resonator mode of the investigated sample types A, B, and C individually by fitting its isolated absorption signature to \cite{probst2015,khalil2012}
\begin{equation}
    S_{11} (f) = A e^{i \phi} e^{-i2\pi  f \tau} \left[1 - \frac{2(Q/|\Qe|)e^{i\theta}}{1+2iQ(f/f_\mathrm{r} - 1)}\right].
    \label{eq:fit}
\end{equation}
Here, the characteristic parameters of the resonance are given by the resonance frequency $f_\mathrm{r}$, as well as the internal, external and total quality factors $\Qi$, $\Qe$, and $Q= 1/(\Qi^{-1}+\Qe^{-1})$, respectively. Furthermore, it takes into account the uncalibrated reflection measurement by introducing a global amplitude factor $A$, an added phase $\phi$, and the delay by the electric length of the cables $\tau$. In addition, the phase $\theta$ compensates for mismatches between input- and output impedance as well as reflection paths in parallel to the resonator. Further details on this so-called 'circle-fit' method can be found in Ref.\,\cite{probst2015}. Figure \ref{fig:circlefit}\,(b) and (c) show an exemplary fit of the SAW resonance highlighted in panel (a) to Eq.\,(\ref{fig:circlefit}). We restrict our analysis to resonances with $|f_\mathrm{r}-f_\mathrm{0}|<5\Delta f_\mathrm{FSR}$ located in the center of the stop band, where we estimate $f_\mathrm{0}$ by taking the average frequency of all visible resonance features.
We also extract the effective length $L_\mathrm{eff}$ of each SAW resonator using $\Delta f_\mathrm{FSR} = v_\mathrm{p}/2L_\mathrm{eff}$, where $v_\mathrm{p}$ is given by the center frequency $f_\mathrm{0}$ of the stop-band. The difference between the designed resonator length $L$ (distance between mirrors) and $L_\mathrm{eff}$ allows us to determine the reflectivity of a single strip $r$ using $r = 2a/(L_\mathrm{eff}-L)$, where $a$ is the width of a single strip \cite{morganSAWFilters}. A summary of the results is presented in Tab.\,\ref{tab:samples}, along with design parameters of each studied device.
In our experiment, we find that the thin film LNO samples exhibit a lower piezo-electrical coupling compared to the bulk system, which we attribute to their reduced thickness compared to the extinction depth of the SAW. This would reduce the external coupling rate of the SAW resonator to the electromagnetic circuit, for which we compensate with the design parameters of the IDT, specifically the number of fingers $N_\mathrm{f}$ and the total width of the SAW resonator $W$ (see also Fig.\,\ref{fig:intro}).\\
\begin{figure}
	 \center{\includegraphics{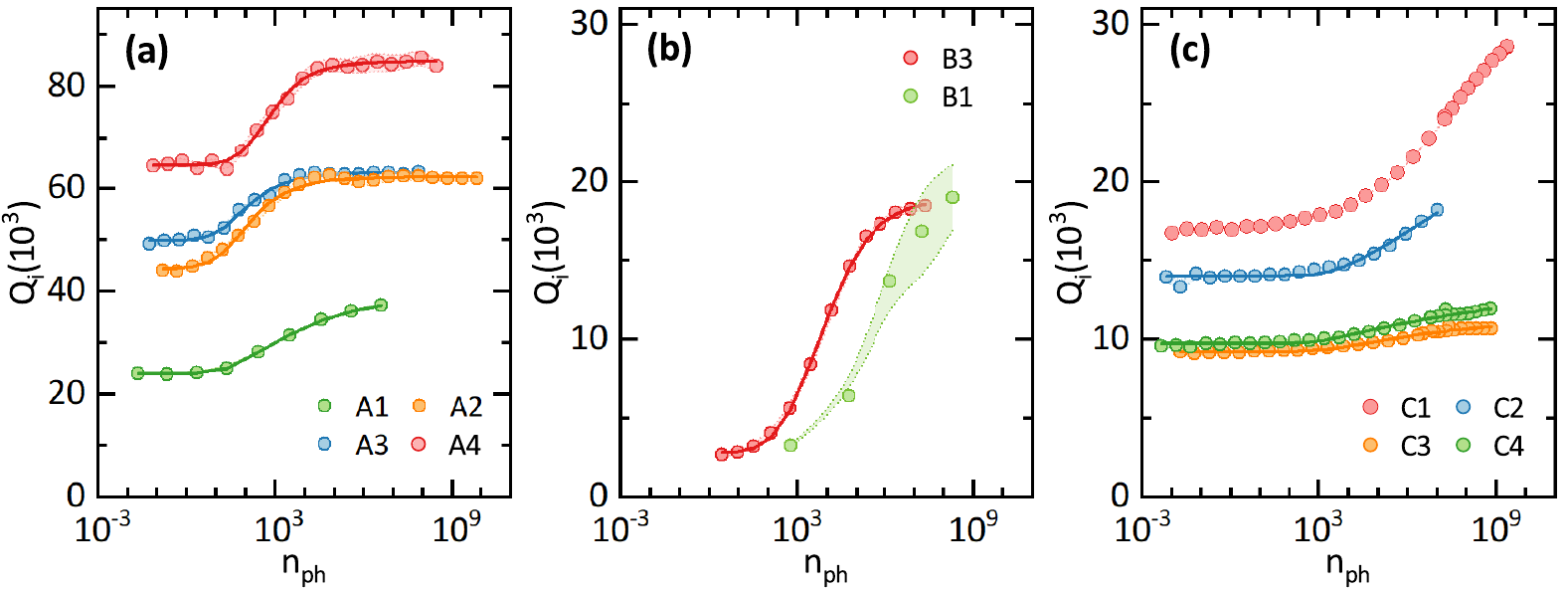}}
	  \caption{\label{fig:qualityfactor} Internal quality factors $\Qi$ as a function of the average phonon number $n_{\textrm{ph}}$ in the resonator mode, extracted by fits to Eq.\,(\ref{eq:fit}) for various samples of types A (bulk LNO), B (LNOI), and C (thin film LNO) (sample details listed in Tab.\,\ref{tab:samples}). For each resonator, the mode with the best ratio of intrinsic quality $\Qi$ factor to its uncertainty $\Qi/\Delta \Qi$ was selected. Uncertainties are derived from the fitting model and are illustrated as shaded areas around the data points. Where possible, the power-dependency of $\Qi$ has been fitted to a TLS based model given by Eq.\,(\ref{eq:Qi_TLS_model}) (solid lines) and the corresponding TLS contribution $Q_\mathrm{i,TLS}$ was extracted and is summarized in Tab.\,\ref{tab:samples}. While the devices on bulk LNO [sample type A, (a)] show a clear saturation behavior for high and low average phonon numbers $n_{\textrm{ph}}$, devices on LNOI [sample type B, (b)] and on LNO thin films [sample type C, (c)] show only partial saturation.}
	\end{figure}
One contribution to the intrinsic losses of solid-state based resonators, which are considered to be of high relevance for the single excitation or low power limit and hence of importance to quantum applications are the losses due to so-called two-level-systems (TLS). For example, in electromagnetic coplanar waveguide resonators, as usually employed in superconducting quantum circuits, TLS losses are commonly identified as the limiting factor for the resonator performance such that $\Qi \approx Q_\mathrm{TLS}$ in the low power limit \cite{goetz2016}. This motivated us to analyze the quality factors of SAW resonators also in the low power or single excitation limit as this regime is the most relevant for quantum acoustic applications. Here, we employ two approaches for their characterization: (i) microwave power-dependent measurements of the quality factors and (ii) temperature dependent characterization of the SAW resonators. The experiments examining the internal quality factor $\Qi$ as function of microwave power are summarized in Fig.\,\ref{fig:qualityfactor}. To put the power levels into perspective, it is useful to convert the incident microwave power at the sample $P$ to an average phonon occupation number $n_\mathrm{ph}$ in the respective resonator mode via 
\begin{equation}
    n_\mathrm{ph} = \frac{4\kappa_\mathrm{e}P}{2\pi\hbar f_\mathrm{r}(\kappa_\mathrm{e}+\kappa_\mathrm{i})^2},
    \label{eq:phonon_nr_conversion}
\end{equation} 
where we have expressed the mode quality factors in terms of loss rates using $\kappa_\mathrm{i,e} = (2\pi f_\mathrm{r}) / Q_\mathrm{i,e}$. \\
We observe an increase in $\Qi$ with increasing drive power for all investigated devices. This behavior is expected where TLS losses dominate and commonly associated with the coupling of the resonator field to a bath of TLS, which are increasingly saturated with increasing excitation number. To quantify this effect, we compare the experimental data to the analytical expression for the loss caused by a TLS bath \cite{pappas2011}:
\begin{equation}
    \frac{1}{\Qi} = \frac{1}{Q_{\mathrm{i,TLS}}}\left(1+\frac{n_\mathrm{ph}}{n_\mathrm{c}}\right)^{- \alpha} + \frac{1}{Q^*} .
    \label{eq:Qi_TLS_model}
\end{equation}
Here, $Q_\mathrm{i,TLS}$ represents the intrinsic loss contribution due to TLS at zero temperature and vanishing photon number $n_\mathrm{ph}$, i.e. no thermal polarization. As the drive power is increased, $n_\mathrm{ph}$ becomes larger, which eventually saturates the TLS and thereby suppresses this loss channel allowing to isolate the TLS losses present in the resonator. The characteristic saturation power for this process expressed as the number of excitations is the critical photon number $n_\mathrm{c}$. In addition, the exponent $\alpha$ accounts for deviations from the standard TLS model ($\alpha = 0.5$ in the standard case). Lastly, $Q^*$ is included to account for all non-TLS contributions. \\
Many of our devices indeed show this characteristic behavior [see Fig.\,\ref{fig:qualityfactor} (a) and (b)] although the impact of TLS losses is less significant compared to their high performance electromagnetic counterparts \cite{mcrae2020}. The corresponding $Q_\mathrm{i,TLS}$ extracted from the fit to Eq.\,(\ref{eq:Qi_TLS_model}) are also summarized in Tab.\,\ref{tab:samples}. In addition, we find for the SAW resonators fabricated using LNO thin films on $\mathrm{Si}$ (sample type C) a qualitatively different behavior, where saturation is not apparent for average phonon numbers up to $10^9$. This set of devices fails an analysis using Eq.\,(\ref{eq:Qi_TLS_model}).
\begin{figure}
  \center{\includegraphics{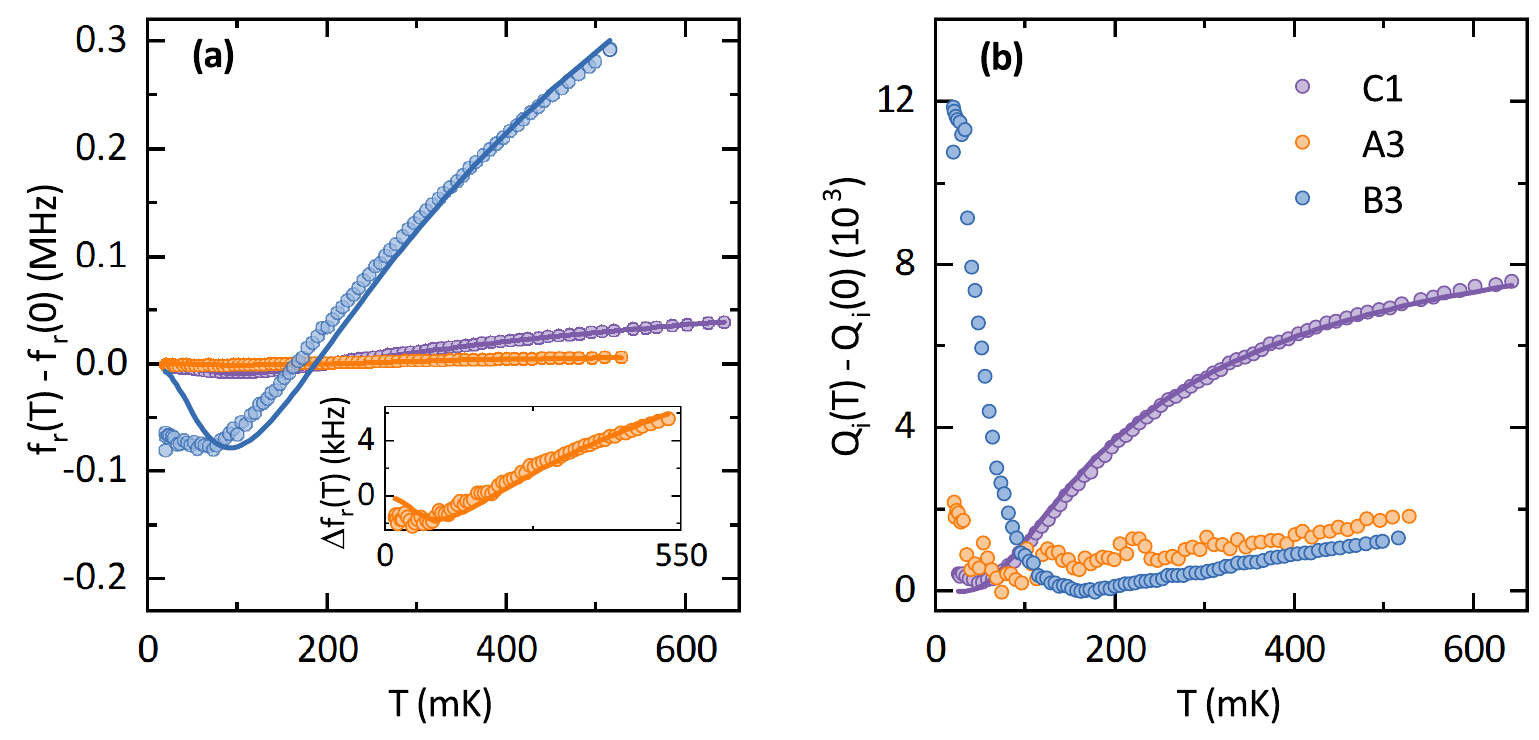}}
  \caption{Experimentally observed change in (a) resonance frequency $f_\mathrm{r} (T)-f_\mathrm{r}(0)$ and (b) intrinsic quality factor $\Qi (T)-\Qi(0)$ as a function of temperature, shown for one selected resonator mode per substrate, labeled as shown in Tab.\,\ref{tab:samples}. The measurements are performed with a constant microwave power corresponding to $n_\mathrm{ph} \approx 10^5$. Solid lines in (a) are fits to Eq.\,(\ref{eq:fitfr}). The inset shows the same experimental data along with the fit to the model for the bulk LNO resonator A3 rescaled to make the much smaller effect visible. In (b) the change in $\Qi$ is plotted relative to the intrinsic quality factor at zero temperature $\Qi(0)$. Fits of $\Qi (T)$ to Eq.\,(\ref{eq:fitQi}) (solid lines) are only possible for the data of resonator C1, while the other two resonators show behavior that is not fully described by the presented TLS model \cite{pappas2011}. Uncertainties as provided by the fit models are smaller than the symbol size.}
  \label{fig:temperatureshift}
\end{figure}
Such behavior is commonly associated with off-resonant TLS, which due to their large detuning from the resonator frequency, do not saturate even at high drive powers \cite{pappas2011}. Temperature dependent measurements allow to provide further insight into these off-resonant TLS as they change the resonator mode frequency $f_\mathrm{r}$ leading to a temperature-dependent frequency shift \cite{pappas2011}
\begin{equation}
    \Delta f_\mathrm{r}(T) = \frac{1}{\pi Q_\mathrm{i, TLS}}\left[\mathrm{Re} \Psi\left(\frac{1}{2}+\frac{h f_\mathrm{r}(T)}{2 \pi i k_\mathrm{B} T}\right)-\ln\left(\frac{h f_\mathrm{r}(T)}{2 \pi k_\mathrm{B} T}\right)\right]f_\mathrm{r}(0),
    \label{eq:fitfr}
\end{equation}
using the digamma function $\Psi$. In addition, they modify the internal Q-factor as
\begin{equation}
    \frac{1}{Q_\mathrm{i}} = \frac{1}{Q_\mathrm{i,TLS}} \tanh\left(\frac{h f_\mathrm{r}(T)}{2 k_\mathrm{B} T}\right)+\frac{1}{Q_\mathrm{res}},
    \label{eq:fitQi}
\end{equation}
where $Q_\mathrm{res}$ describes non-TLS contributions.\\
Figure \ref{fig:temperatureshift}\,(a) and (b) show the temperature dependent $f_\mathrm{r}$ and $\Qi$ measurements for one device of each of the investigated sample types, performed at constant microwave powers corresponding to $n_\mathrm{ph} \approx 10^5$. Pronounced changes in $f_\mathrm{r}$ as well as $\Qi$ are observed for devices C1 and B3, i.e.\ on LNO thin film sample types, while device A3 fabricated on bulk LNO shows significantly weaker temperature dependence. Therefore, it can be assumed that thermal TLS do not contribute as strongly to the losses in bulk LNO. This is expected from the observation made from Fig.\,\ref{fig:qualityfactor} (a), where a clear saturation of $\Qi$ as a function of power suggests that the most relevant loss mechanisms for SAW resonators on bulk LNO are well described by Eq.\,(\ref{eq:Qi_TLS_model}) even without considering thermally excited TLS. It is notable, however, that our data of resonator B3 deviates significantly from the behavior predicted by the off-resonant TLS model for temperatures below $\SI{100}{mK}$. An unexpected and large increase in $\Qi$ is observed, as well as a constant $f_\mathrm{r}$ where a characteristic local minimum is expected. We believe that this effect can be attributed to a special class of resonant TLS, which are not considered by either of the applied theoretical models: Due to the low overall quality factor of the LNOI resonator B3, it is possible that the characteristic coherence times of some resonant TLS become comparable to the loss rate of the resonator itself and the TLS can no longer contribute a significant additional loss \cite{place2022}. This effect would become visible for temperatures where thermal excitation of these TLS is no longer possible (i.e.\ below $\SI{100}{mK}$ for TLS frequencies of $f_\mathrm{TLS}\approx \SI{4}{GHz}$), which agrees well with our experimental observation. A full description of this behavior would require a more extensive theory considering coherent exchange of excitations with TLS, which would exceed the scope of this work. Despite this discrepancy, by fitting the temperature dependence of devices A3, B3 and C1 to Eq.\,(\ref{eq:fitfr}), we extract equivalent quality factors $Q_\mathrm{TLS,T}^\mathrm{A3} = (155.6 \pm 1.1) \times 10^{3}$, $Q_\mathrm{TLS,T}^\mathrm{B3} = (3.7 \pm 0.1) \times 10^{3} $ and $Q_\mathrm{TLS,T}^\mathrm{C1} = (36.2 \pm 0.2) \times 10^{3}$ for the thermal TLS contribution. Further, a fit to Eq.\,(\ref{eq:fitQi}) for device C1 results in $Q_\mathrm{TLS,T}^\mathrm{C1} = (27.8 \pm 0.4) \times 10^{3}$. These extracted quality factors show excellent agreement with the previous analysis of power-dependency, suggesting that internal losses of our devices are in fact limited by TLS and can be appropriately modelled using the currently employed formulas. 

The extracted quality factors along with geometric parameters for all of the investigated devices are summarized in Table \ref{tab:samples}.
In particular, we observe $Q_\mathrm{TLS,P} \approx Q_\mathrm{TLS,T} \gg \Qi$ for devices fabricated on bulk LNO, suggesting that TLS are not dominating the losses of these devices and other mechanisms, e.g.\ acoustic losses to bulk waves, play a larger role. For bulk devices, this conjecture is supported by the large dependence of $\Qi$ on geometric parameters (i.e.\ the aperture $W$) (cf.\ samples A1-A4 in Tab.\,\ref{tab:samples}). Conversely, for the investigated devices fabricated on LNO thin film samples, we observe $Q_\mathrm{TLS,P} \approx Q_\mathrm{TLS,T} \approx \Qi$, which is consistent with TLS currently limiting the performance of these devices. We attribute this difference in loss behavior to the additional interfaces associated with
the multi-layer structure. Studies from cQED show that interfaces and, in particular, surface oxides can lead to large TLS losses in coplanar waveguide resonators \cite{mcrae2020}. For LNOI samples, this is especially expected due to the introduction of the silicon dioxide buffer layer, which is known to be a lossy dielectric. We also suspect that during the wafer-bonding process used to fuse thin film LNO to silicon, a non-negligible thermal oxide layer is formed in between the two layers, producing additional interfaces and potential TLS, warranting a more detailed study of the LNO-Si interface. It is worth noting that there are experiments performed at room temperature which have found SAW devices on LNOI films to exhibit higher quality factors compared to devices on bulk LNO \cite{kimura2019}, in contrast to the results of our study. This underscores the dominating effect of TLS-related losses for devices operating in the quantum regime. This is in clear contrast to devices operated at room temperature, where TLS are expected to be fully saturated and hence TLS losses are usually negligible.

\section{Conclusion}
	In this work, we have investigated surface acoustic wave resonators fabricated on thin film lithium niobate and lithium niobate on insulator (LNOI) samples and compared their performance at millikelvin temperatures and in the quantum regime to reference devices fabricated on the well established piezo-electric lithium niobate (bulk). As a figure of merit, we focus on the internal quality factor $\Qi$ of the resonators, which can be used as an indication for the losses that might limit quantum applications of the SAW devices. We find that internal quality factors of the devices fabricated on multi-layer substrates are generally lower than of those fabricated on bulk LNO, but remain roughly in the same order of magnitude. Subsequently, we used power- and temperature-dependent measurements to gain further insight into the underlying loss mechanisms present in the different substrates, with particular focus on the contribution of two-level systems (TLS), which are often limiting the performance of electromagnetic resonators used in circuit electrodynamics. Using two different models based on thermal and non-thermal TLS, we are able to describe our experimental result and extract the contribution of these TLS to the total losses of the devices. In summary, even with the slightly increased loss rates compared to our bulk LNO devices, SAW resonators on thin film LNO demonstrate performance in the single-phonon regime that is comparable to devices on bulk LNO, which are successfully used in quantum acoustic circuits \cite{satzinger2018,bienfait2019}. Moreover, potential advantages in scalability and versatility by avoiding flip-chip assemblies might favor the use of thin film SAW resonators over established bulk devices for particular applications.
\ack
We acknowledge funding from the European Union's Horizon 2020 Research and Innovation Programme under grant agreement No 736943 and from the Deutsche Forschungsgemeinschaft (DFG, German Research Foundation) under Germany’s Excellence Strategy—EXC-2111-390814868. Further, this work was supported in part by the Horizon Europe 2021-2027 Framework Programme under the grant agreement No 101080143 (SuperMeQ).
\section*{References}
\providecommand{\newblock}{}

\end{document}